\documentclass[useAMS,usenatbib]{mn2e}
\usepackage{epsfig}
\usepackage{graphicx}

\begin{document}
\voffset=-0.5 in

\title[Hyperflares of SGRs and mERBs]{Hyperflares of SGRs as an engine for millisecond extragalactic radio bursts}

\author[S.B. Popov, K.A.Postnov]{S.B. Popov$^{1}$, K.A.Postnov$^{1,2}$
\thanks{E-mail: polar@sai.msu.ru(SBP); pk@sai.msu.ru(KAP)}\\
$^1${Sternberg Astronomical Institute, Universitetski pr. 13, Moscow
119992, Russia} \\
$^2${Institute of Astronomy and Astrophysics, Kepler Center for Astro and Particle Physics, University of Tuebingen, Germany} }

\date{Accepted ......  Received ......; in original form ......
}


\maketitle

\begin{abstract}
 We propose that the strong millisecond extragalactic radio burst (mERB) discovered by Lorimer et al. (2007) may
 be related to a hyperflare from an extragalactic soft gamma-ray repeater. The expected rate of such hyperflares, $\sim$~20~-~100~d$^{-1}$~Gpc$^{-3}$, is in good correspondence with the value estimated by Lorimer et al. The possible mechanism of radio emission can be related to the tearing mode instability 
 in the magnetar magnetosphere as discussed by Lyutikov (2002), and can produce 
 the radio flux corresponding to the observed $\sim$~30 Jy from the mERB using a simple scaling of the burst energy.
\end{abstract}

\begin{keywords}
stars: neutron --- gamma-rays: bursts --- pulsars: general
\end{keywords}


\section{Introduction}
\label{intro}

Progress in observational technique in radio astronomy made it possible to detect single millisecond scale bursts \citep{cm2003}. Recently
\cite{merb2007} reported a serendipitous discovery of a strong non-thermal
millisecond radio burst with peculiar properties.
The radio flux at 1.4 GHz is $30\pm 10$ Jy, the duration of the event is
shorter than 5~msec and its dispersion measure (DM) is 375 cm$^{-3}$pc
suggesting the extragalactic nature of the source at a distance of  $<$~1~Gpc.
No host galaxy up to 18 $B$-magnitude has been found
implying a distance limit of $>$600~Mpc for a Milky Way-like host.
Such a distance implies a radio energy release  in the burst of $\sim 10^{40}$ ergs
with a brightness temperature of  $\sim 10^{34}$ K.
The total rate of such bursts is estimated to be around 90~d$^{-1}$~Gpc$^{-3}$,
which is much lower than the core-collapse supernova (SN) rate --
$\sim$~1000~d$^{-1}$~Gpc$^{-3}$,
but well in excess of the gamma-ray burst (GRB) rate --
$\sim$~4~d$^{-1}$~Gpc$^{-3}$, and the expected rate
of binary neutron star coalescences  $\sim$~2~d$^{-1}$~Gpc$^{-3}$.

The short time scale means the origin in a compact region $<1500$ km (in
the non-relativistic case), suggesting
neutron stars as the most probable sources of mERBs.
\cite{merb2007} discuss several possible known sources of the
strong millisecond radio bursts,
including rotating radio transients (RRATs) and giant pulses from radio pulsars, but none
of them appear to be energetic enough.

However, if the transient is related to the energy release by a compact magnetized object
characterized by the rotational period $P$ and the dipole magnetic moment $\mu$, 
the following scaling considerations may be appropriate 
to provide the observed $10^{40}$ ergs in the radio burst. For example, 
the giant radio pulses in the Crab pulsar with a peak luminosity of 4~kJy~kpc$^{-2}$ can
be observed in a search like the one performed by \cite{merb2007}
from distancies up to 100 kpc. Although the precise mechanism
of the giant pulses is still unknown, we can assume that at least this amount
of energy can be provided by some mechanism in the radio band. Then simple scaling of the Crab 
rotational power to 500 Mpc yields 
$2.5\times 10^7$ as much a rotational energy release, i.e. $2.5\times 10^{45}$
erg~s$^{-1}$.
 As $\dot E_{rot}\propto \mu^2P^{-4}$, this power 
is attainable for a msec spin period and
the magnetic field an order of magnitude higher than that of the Crab pulsar. These physical conditions can
in principle be realized during the birth of a ms magnetar \cite{usov1992}
or at the late phases of coalescence of binary
neutron stars with magnetic fields, as was considered by \cite{lp1996} (see
also \citealt{v1996} and \citealt{hl2001}).

Unfortunately, binary neutron star coalescences,  
even considering the high uncertainty in the
current estimates of their galactic rate \citep{py2006}, 
are expected to happen almost a hundred times less frequently than the reported rate of mERBs.
In addition, they are thought to be accompanied by short gamma-ray bursts (\citealt{b1984}, see the review in \citealt{n2007}).
The fact that none was
detected at the time of this radio burst can be due to gamma-ray beaming off-set or
intrinsic weakness, but non-detection of msec periodicity may be
a stringer constraint. Observations of the simultaneous 
gravitational wave burts and gamma-ray burst will be decisive for the
binary NS nature of the mERB.

Magnetars appear more preferential. Their estimated rate of birth (roughly, $10^{-3}$~yr$^{-1}$ per galaxy;
\citealt{wt2006}, and may be even higher; see P.~Wood, 2007, talk at the conference "40-years of pulsars") can be consistent with 
that of mERBs. The non-thermal coherent radio emission can also be produced 
at their birth by instabilities in relativistic 
plasma which can be created by the Poynting flux outflow \cite{usov1994}.

However, both binary neutron star mergings nor magnetar birth have not been
firmly observed so far.
Here we propose to discuss another possibility 
that the strong mERB could be associated with
hyperflares of extragalactic soft gamma-ray repeaters.

\section{mERBs from SGRs}

Soft gamma-ray repeaters (SGRs) show a very complicated outburst activity
in hard rays (\citealt{h1999}, for a review see \citealt{wt2006}), but the most
pronounced are hyperflares (HFs)
like the event on Dec. 27, 2004  from SGR 1806-20
(\citealt{betal2005}) with an energy release of $>$~a few~10$^{46}$~ergs
(for observational details see, for example,
\citealt{h2005, petal2005}).
Recently, several candidates for extragalactic HFs have been proposed (\citealt{f2006, o2006}, see aslo \citealt{g2007, c2006}).

Is it possible that the mERB is related to an extragalactic HF?
To address this question, 
we first note that the inferred rate of mERBs
is similar to the rate of HFs of SGRs
obtained from searches for SGR flares in close-by galaxies and the Virgo cluster.
\cite{ps2006} argue that the rate of HFs is 
$\sim$~10$^{-3}$~yr$^{-1}$ per a Milky Way-like galaxy.
\cite{lgg2005}
give slightly less stringent limit for the rate of HFs: $<$~1/130~yr$^{-1}$. Considering low statistics, however, 
here we use a more conservative estimate 0.001-0.003 per year per a Milky Way-like galaxy.
This is $\sim$~10~-~50 times smaller than the galactic rate of SN, so
from the SN rate $\sim$~1000~d$^{-1}$~Gpc$^{-3}$ we estimate the
expected rate of SGR HFs to be $\sim$~20~-~100~d$^{-1}$~Gpc$^{-3}$
in good correspondence
with the rate of mERBs obtained by \cite{merb2007}.

The possible mechanism of a prompt intense radio burst from SGR flares was discussed by
\cite{l2002} (see also \citealt{l2006}).
He proposed that a $\sim$~10~msec radio burst can be generated in the magnetar magnetosphere 
due to the tearing mode instability, following the similarity between solar
flares and SGR bursts (in solar flares radio bursts accompany X-ray flares). For galactic SGRs
with X-ray luminosity $10^{36}$~-~$10^{39}$~erg~s$^{-1}$ and 10~kpc
distance he estimated a possible radio flux at $\sim$~1~GHz
of about 1~--~1000~Jy. Guided by these values, for a HF with peak gamma-ray luminosity of
$L=10^{47}$~erg~s$^{-1}$ (as observed in the Dec. 27 2004 HF from SGR 1806-20)
and 600~Mpc we obtain a radio flux of $\sim$~30~Jy, in correspondence with observations by \cite{merb2007}.

The millisecond time scale of the mERB is consistent with an event in a magnetar magnetosphere.
For example, the light curve of the Dec. 27, 2004 event (see Fig.~1b in \citealt{petal2005})
shows a few features with the duration of a few milliseconds at the initial stage.
The raising part of the main spike of the Dec 27, 2004 HF is also about 5~msec.
The time scale of the radio burst produced by the discussed mechanism 
can be much smaller than 5~msec,
since the crossing time of Alfven waves in internal parts of the NS magnetosphere
is just $t_\mathrm{A}\sim R_\mathrm{NS}/c\sim 30 \mu$sec (see discussion in \citealt{l2006}).
We note here that \cite{merb2007}
actually stress that the observed burst could indeed be
much shorter than 5 msec.

In addition to strong HF, SGRs show less intensive and more frequent 
giant flares (GFs), like those on March 5, 1979 from SGR 0526-66 and Aug. 27, 1998 from SGR 1900+14. 
Their galactic rate is estimated to be about 0.05~-~0.02~yr$^{-1}$, 
similar to the SN rate (Woods \& Thompson (2006)\nocite{wt2006}.
Because of low statistics
it is unknown if GFs 
and HFs from SGRs form continuous luminosity distribution. If they do, then mERB
energy distribution should follow the same power law 
$dN/dE\sim E^{-\gamma}$ with the index $\gamma \approx 1.6$~-~$1.7$ \citep{g1999}.
But then, as noted by \cite{merb2007}
one would expect to see more weaker mERBs,
which is apparently not the case.
If one assumes that these events belong to different classes without bursts of intermediate energies,
it is possible to explain why the first detected mERB was about two orders of magnitude above the threshold.
If GFs can also produce mRBs with energies scaling similar to X-ray~-~soft-$\gamma$~bursts,
as proposed by \cite{l2002},
then they are too dim (or rare, if they come from much smaller volume corresponding to smaller distancies) to be detected in a search
like the one performed by \cite{merb2007}.
If the HF which produced the mERB was followed by usual
(weak) bursts with energies $<$~10$^{41}$~erg~s$^{-1}$, as it is assumed to be typical for GFs and the HFs, then these events
would produce too dim radio bursts (0.03 mJy according to the scaling suggested by \citealt{l2002})
to be detected.

\section{Discussion}

SGRs are assumed to be young ($<$~10$^4$~yrs) neutron stars, so 
extragalactic flares are expected to be related 
to galaxies with high star formation rate \citep{ps2006}.
Potentially, this can give a hint about possible properties
of the host galaxy of the mERB discovered by \cite{merb2007}.
For example, the possible host galaxy can be rather dim in the optical
due to dust; not necessarily it should be a Milky Way-like galaxy,
it can be a smaller irregular galaxy with active star formation.
In this case it can lie closer than the 600~Mpc distance as inferred from the optical limit.
Intrinsic DM can also be higher in star forming galaxies.
So it would be interesting to use the Spitzer space telescope to search for the host
galaxy of this event in the infrared.

Unfortunately, there is no much hope to find many mERBs in dedicated searches 
for activity of extragalactic SGRs, as HFs are very rare events. 
Radio monitoring of galaxies even with extremely
high rate of star formation appears to be not very promising in the sence of looking for more mERBs.
For example, even in the ``SN factories'' discussed in Popov, Stern (2006)
with the star formation rate about two orders of magnitude as high as in the Milky Way,
we can expect to have only few extragalactic giant flares per year (from their distances of tens Mpc 
the scaling discussed by Lyutikov 2002 would yield detectable GFs-related radio bursts).

Nevertheless, when performing archive searches for more mERB candidates
it can be useful to check first directions with large integrated star
formation rate along the line of sight. To find more mERBs related to SGR
HFs this could be more promising than the direct search in programs like 
STARE \citep{ketal1998} and FLIRT (\citealt{balsano}), because mERBs can have no counterparts in other wavelengths.

To our knowledge, nobody was able to test directly the prediction by \cite{l2002} by 
observing galactic SGRs and/or AXPs exactly during bursts in radio 
with msec time resolution. Recently, some observations of AXPs have been reported \citep{bur2006, chk2007}. These authors put strong limits on the integrated pulsed emission and RRAT-like bursts of several AXPs, however, nothing can be said about mRB during X/$\gamma$~bursts.

Here we note that the absence of high energy bursts
during RRAT flares
suggests a different mechanism from the one discussed by \cite{l2002} for radio flares from SGRs. Observed radio fluxes of RRATs are about 0.1~-~4~Jy \citep{ml2006}, which in the model from \citep{l2002}corresponds to X/$\gamma$-ray luminosities about
10$^{35}$~-~10$^{36}$~erg~s$^{-1}$ for the estimated RRATs distancies $\sim$~2~-~7~kpc. For X-ray bursts several hundred milliseconds long this is bright enough to be detected by RXTE taking into account that bursts are frequent and regular. Also, stringent limits have been put on X-ray bursts from one of the RRAT J1819-1458 observed by Chandra \citep{ml2007}. However, if X-ray bursts are very short, then fluence can be low, below the detection threshold.

Lorimer et al. (2007) note that no gamma-ray burst was detected at the moment of the mERB
(August 24, 2001). In the case of a HF from the distance $\sim$~600~Mpc
it is not surprising. For example, BATSE could detect
events like the Dec. 27, 2004 flare only from a distance of
$\sim$~30~-~40~Mpc \citep{h2005, ps2006}.
SWIFT can detect HFs (similar to Dec. 27, 2004) from larger distances
($\sim$~70~Mpc, \citealt{h2005}),
which is still much lower than 600~Mpc.

After the HF of Dec. 27, 2004  (as also after the Aug. 27, 1998 one),
a weak radio source have been detected (see, for example,
\citealt{g2005,g2006} and references therein).
However, it is impossible to register such a dim object from the distance $\sim$~600~Mpc.

%

 Some other exotic possibilities, like deconfinement inside a compact object
leading to a quark star formation and complete reconfiguration of the magnetic field, can be discussed.
Most of them are also expected to be accompanied by a SN-like or/and a GRB-like event.

We conclude that the rate of HFs of SGRs of several tens per day within 1 Gpc volume
is similar to the inferred rate of mERBs. The time scale,
the absence of counterparts at other wavelengths, as well as the non-observation of
the host galaxy up to the 18th $B$-magnitude,
are consistent with the HF hypothesis. 
The physical mechanism of the radio burst can be related 
to the tearing mode instability, as proposed by \cite{l2002}.
New observations are decisive to check if extragalactic binary neutron star mergings or hyperflares from SGRs underly mERBs.

\section*{Acknowledgments}
The work of KAP is partially supported by DAAD grant A/07/09400.
SBP acknowledges  the INTAS fellowship (in part with the Observatory of Cagliari), and thanks the Observatory of Cagliari and 
the University of Cagliari for hospitality, and colleagues from these institutions for stimulating discussions. We thank Maxim Lyutikov for useful comments on the manuscript.

\end{document}